\documentclass[aps,prl,twocolumn,showpacs]{revtex4-1}
\usepackage{graphicx}
\usepackage{bm,braket}
\begin{document}
\title{Investigation of the Superconducting Gap Structure in  SrFe$_2$(As$_{0.7}$P$_{0.3}$)$_2$ by \\Magnetic Penetration Depth and Flux Flow Resistivity Analysis }

\author{Hideyuki Takahashi$^{1,3}$, Tatsunori Okada$^{1,3}$, Yoshinori Imai$^{1,3}$, Kentaro Kitagawa$^{2}$,\\ Kazuyuki Matsubayashi$^{2}$, Yoshiya Uwatoko$^{2}$, and Atsutaka Maeda$^{1,3}$}

\affiliation{
$^{1}$Department of Basic Science, the University of Tokyo, 3-8-1 Komaba, Meguro-ku, Tokyo 153-8902, Japan\\
$^{2}$Institute for Solid State Physics, University of Tokyo, Kashiwa, Chiba 277-8581, Japan\\
$^{3}$IRON-SEA, JST, Japan}

\date{\today}

\begin{abstract}
We measured the microwave surface impedances and obtained the superfluid density and flux flow resistivity in single crystals of a phosphor-doped iron-based superconductor SrFe$_2$(As$_{1-x}$P$_{x}$)$_2$ single crystals ($x=0.30$, $T_c=25\ \mathrm{K}$). At low temperatures, the superfluid density, $n_s (T)/n_s(0)$, obeys a power law, $n_s (T)/n_s (0)=1-C(T/T_c)^n$, with a fractional exponent of $n=1.5$-1.6. The flux flow resistivity was significantly enhanced at low magnetic fields. These features are consistent with the presences of both a gap with line nodes and nodeless gaps with a deep minimum.
The remarkable difference observed in the superconducting gap structure between SrFe$_2$(As$_{1-x}$P$_{x}$)$_2$ and BaFe$_2$(As$_{1-x}$P$_{x}$)$_2$ in our experiments is important for clarifying the mechanism of iron-based superconductivity.
\end{abstract}

%\pacs{74.25.nn, 74.20.Rp, 74.25.fc, 74.62.-c}

\maketitle
The symmetry of the superconducting (SC) order parameter is closely related to the pairing interaction and is an important problem for iron-based superconductors~\cite{Kamihara08}. While nodeless $s\pm$-wave symmetry mediated by antiferromagnetic (AF) spin fluctuation has been considered to be a very likely candidate~\cite{PhysRevLett.101.057003, PhysRevLett.101.087004}, it has been revealed that many compounds have line nodes in their SC gap function.
P-substituted systems are especially interesting in terms of the effect of the local structure parameter on the SC gap function.
Kuroki \textit{et al}. suggested that line nodes appear in the SC gap function when the pnictogen height from the Fe-As plane decreases~\cite{PhysRevB.79.224511,JPSJ.80.013710}. 
One can experimentally control the pnictogen height and induce superconductivity through the isovalent substitution of As by P in the 122 system.
BaFe$_2$(As$_{0.67}$P$_{0.33}$)$_2$ ($T_c\sim 30\ \mathrm{K}$) is one of the most intensively studied P-substituted compounds~\cite{PhysRevB.81.184519}. 
London penetration depth measurement~\cite{PhysRevB.81.220501} provides an evidence of line nodes as other P-substituted iron-based SCs, such as LaFePO~\cite{PhysRevLett.102.147001} and LiFeP~\cite{PhysRevLett.108.047003}, and the presence of closed nodal loops in the flat region of the electron Fermi surface was proposed based on an angle-resolved thermal conductivity measurement~\cite{PhysRevB.84.060507}. 
On the other hand, it has been theoretically suggested that P-substitution warps the outer hole Fermi surface and weakens the nesting between the hole and electron Fermi surfaces. This action reduces the two-dimensionality of the spin fluctuation, which results in the appearance of three-dimensional nodes on the outer hole sheet~\cite{JPSJ.80.013710}.
Therefore, there is no consensus on the detailed structure of the nodes.
Another important consideration is to determine why the P-substituted Ba122 compounds have rather high $T_c$ values, even in the presence of nodes. Therefore, it is of critical importance to investigate the SC gap structure in other P-doped systems.

In this paper, we focus on a different 122 system, namely Sr122 system. 
The parent compound SrFe$_2$As$_2$ exhibits a maximum transition temperature of $T_c^{\mathrm{max}}=32\ \mathrm{K}$ at high pressures ($P=5\ \mathrm{GPa}$)~\cite{JPSJ.78.073706}.
Around the optimum pressure, nuclear magnetic resonance measurements have revealed the SC/AF hybrid state and the development of spin-fluctuation toward $T_c$~\cite{PhysRevLett.103.257002}.
The chemical substitution is also effective in inducing superconductivity~\cite{PhysRevLett.101.107007,PhysRevLett.101.207004,0953-8984-22-12-125702,JPSJ.79.095002}.
We measured the microwave surface impedances of SrFe$_2$(As$_{1-x}$P$_x$)$_2$ single crystals ($x=0.30$, $T_c=25\ \mathrm{K}$) to investigate the SC gap structure. 
We found that the low-temperature superfluid density, $n_s$, and the flux flow resistivity, $\rho_f$, at low fields are consistent with the presence of a SC gap that has line nodes. 
However, remarkable differences from BaFe$_2$(As$_{0.67}$P$_{0.33}$)$_2$ were also found in the details of the SC gap, which is an important for clarifying the mechanism of iron-based superconductivity.

%%%%%%%%%%%%%%%%%%%%%%%%%%%%%%%
%%%%%%%%%%FIgure
%%%%%%%%%%%%%%%%%%%%%%%%%%%%%%%
\begin{figure}[tb]
\begin{center}
\includegraphics[width=8.7cm]{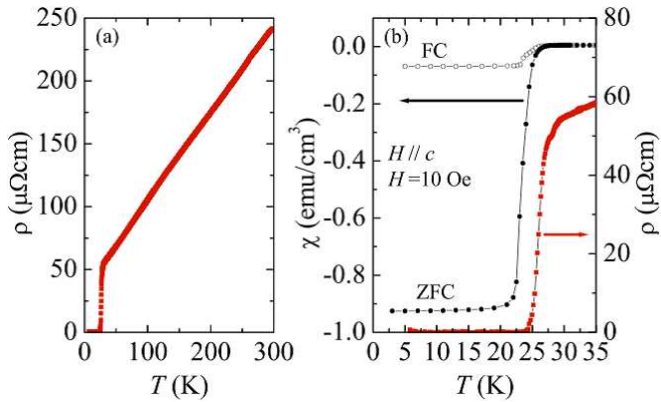}
\caption{(color online) (a) The temperature dependence of the dc resistivity of a SrFe$_{2}$(As$_{0.7}$P$_{0.3}$)$_2$ single crystal. (b) The dc resistivity and the dc magnetic susceptibility with both zero-field cooled (ZFC) and field cooled (FC) conditions at low temperatures.}
\label{transport}
\end{center}
\vspace{-6mm}
\end{figure}
%%%%%%%%%%%%%%%%%%%%%%%%%%%%%%%

Single crystals of SrFe$_2$(As$_{1-x}$P$_x$)$_2$ were grown using the SrAs self-flux method. 
Energy-dispersive x-ray analysis revealed that P substituted 30\ \% of the As sites, which corresponds to $x=0.30$.
Figure~\ref{transport} shows dc resistivity and dc magnetic susceptibility as functions of temperature of our SrFe$_2$(As$_{0.7}$P$_{0.3}$)$_2$ single crystals.
The dc resistivity exhibits a $T$-linear dependence over a wide temperature range, which is characteristic of the materials near the quantum critical point~\cite{PhysRevB.81.184519,Kobayashi,Nature.433.226}. 
The SC transition temperature determined from the zero resistivity temperature is $T_c=25~\mathrm{K}$. The sharp transition and perfect shielding indicate that the crystals are of high quality.

The surface impedance $Z_s=R_s-iX_s$, where $R_s$ is the surface resistance and $X_s$ is the surface reactance, was measured using a cavity perturbation technique.
We used two kinds of cylindrical oxygen-free Cu cavity resonators, operating in the TE$_{011}$ mode at 19~GHz and 44~GHz, which had quality factors, $Q$, of $\sim60000$ (19~GHz) and 26000 (44~GHz), respectively. 
A piece of the crystal was mounted on a sapphire rod and was placed in the antinode of the microwave magnetic field $H_{\omega}$. The shielding current flows in the $ab$ planes because $H_{\omega}$ is parallel to the $c$-axis. In this technique, one measures the changes of $Q$ and the resonant frequency, $f$, of the resonator that result from the introduction of a sufficiently small sample. The shifts in the inverse of $Q$ and $f$ are proportional to $R_s$ and $X_s$, respectively. The absolute values of $Z_s$ were obtained by assuming the Hagen-Rubens limit ($\omega\tau\ll 1$), $R_s=X_s=(\mu_0 \omega \rho/2)^{\frac{1}{2}}=\frac{1}{2}\mu_0 \omega\delta $, above the $T_c$ ($\omega=2\pi f$ is the angular frequency, $\tau$ is the quasiparticle scattering time, $\mu_0$ is the permeability in vacuum, and $\delta$ is the skin depth).
 
 %%%%%%%%%%%%%%%%%%%%%%%%%%%%%%%
%%%%%%%%%%FIgure
%%%%%%%%%%%%%%%%%%%%%%%%%%%%%%%
\begin{figure}[tb]
\includegraphics[width=8.6cm]{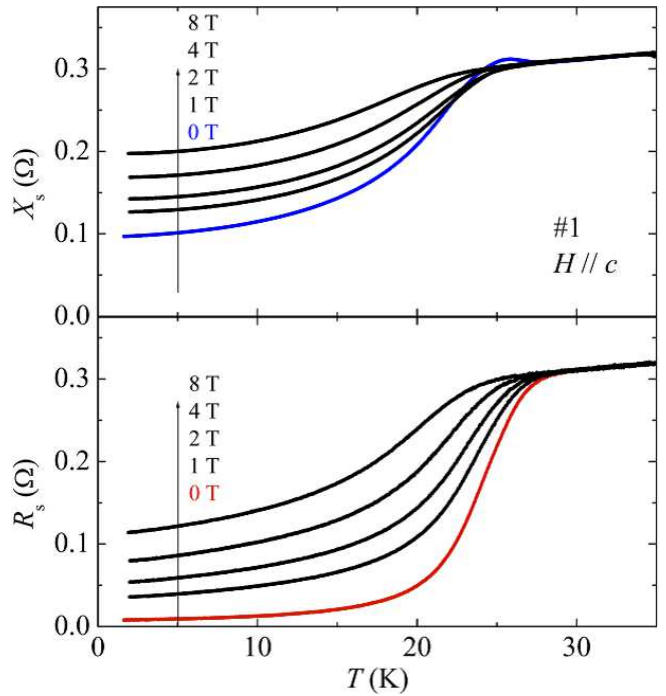}
\vspace{-3mm}
\caption{(color online) The temperature dependence of the microwave surface impedances of SrFe$_{2}$(As$_{0.7}$P$_{0.3}$)$_2$ single crystals at 44~$\mathrm{GHz}$. The hump appears at $T_c$ in the $X_s$ data because the characteristic field-penetration length changes from $\delta$ (in the normal state) to $\lambda$ ($>2\delta$ in the SC state) within a narrow temperature range.}
\label{Zs}
\vspace{-5mm}
\end{figure}
%%%%%%%%%%%%%%%%%%%%%%%%%%%%%%%

In the SC state in the low temperature limit, the surface reactance is proportional to the London penetration depth ($X_s =\mu\omega\lambda_L$). 
We can obtain the superfluid density $n_s (T) /n_s (0)=\lambda_L^2(0)/\lambda_L^2(T)$ via the London equation $\lambda_L^{-2}=\mu_0 n_s e^2/m^*$, which will provode information about the SC gap structure, particularly the presence or the absence of nodes in the gap.
For example, at low temperatures, for superconductors with line nodes, $n_s (T) /n_s (0)$ is expressed as
\begin{equation}
\frac{n_s (T)}{n_s(0)}=1-\alpha T,
\end{equation}
where $\alpha$ is a constant, and it is expressed for nodeless superconductor as
\begin{equation}
\frac{n_s (T)}{n_s(0)}=1-\sqrt{\frac{2\pi\Delta_{\mathrm{min}}}{k_B T}}\exp(-\frac{\Delta_{\mathrm{min}}}{k_B T}),
\end{equation}
where $\Delta_{\mathrm{min}}$ is the minimum amplitude of the SC gap.
We note that the discussion on the temperature dependence of $n_s$ is essentially the same as that on $\delta\lambda_L(T)=\lambda_L(T)-\lambda_L(0)$ at low temperatures because $n_s (T) /n_s (0)\simeq 1-2\delta\lambda_L/\lambda_L (0)$ when $\delta\lambda_L/\lambda_L(0)\ll 1$.

%%%%%%%%%%%%%%%%%%%%%%%%%%%%%%%
%%%%%%%%%%FIgure
%%%%%%%%%%%%%%%%%%%%%%%%%%%%%%%
\begin{figure}[tb]
\begin{center}
\includegraphics[width=8.7cm]{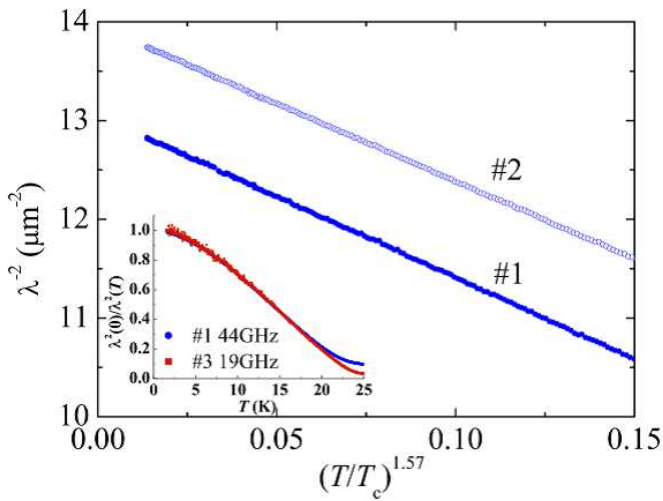}
\caption{(color online) The magnetic penetration depth of SrFe$_2$(As$_{0.7}$P$_{0.3}$)$_2$ single crystals at low temperature measured at $44~\mathrm{GHz}$. The data of crystal \#2 are shifted vertically $+1\mathrm{\mu m^{-2}}$ for clarity. The inset shows the temperature dependence of the superfluid densities $\lambda_L^2(0)/\lambda_L^2(T)=n_s(T)/n_s(0)$ measured at different frequencies. }
\label{lam}
\end{center}
\vspace{-6mm}
\end{figure}
%%%%%%%%%%%%%%%%%%%%%%%%%%%%%%%
Figure \ref{Zs} shows the temperature dependence of microwave surface impedances of a SrFe$_2$(As$_{0.7}$P$_{0.3}$)$_2$ sample with dimensions of $0.5\times0.5\times0.1\ \mathrm{mm^3}$.  
The Hagen-Rubens relation holds well; therefore, we were able to determine the absolute value of the London penetration depth with good reproducibility. London penetration depth at $0~\mathrm{K}$ is approximately $270~\mathrm{nm}$. This value is comparable to those of other optimally doped 122 compounds~\cite{PhysRevB.81.220501,0034-4885-74-12-124505}. 
The superfluid density measured at two different frequencies (See the inset in Fig.~\ref{lam}) are in very good agreement, which indicates a good reliability of the absolute value of $\lambda_L$.

The temperature dependence of the superfluid densities, $n_s (T) /n_s (0)$, up to 0.3$T_c$ is shown in Fig.~\ref{lam}.
Unlike BaFe$_2$(As$_{0.67}$P$_{0.33}$)$_2$, a $T$-linear dependence was not observed.
For all samples, the data were well fitted by a power law, $n_s (T)/n_s (0)=1-C(T/T_c)^n$, with a fractional exponent, $n\simeq 1.6$, and $n$ did not depend on the fitting temperature range.
The power law has been observed in many iron-based compounds~\cite{0034-4885-74-12-124505}. However, the values of the exponents are 2 or greater in most cases. 
For the case of $s\pm$-wave superconductors, the intrinsic exponential temperature dependence of $\lambda_L(T)$ changes to exhibit a power-law behavior when nonmagnetic disorder is introduced~\cite{PhysRevB.79.140507}. As disorder increases, the exponent decreases from $n>3$ to $n\simeq2$~\cite{PhysRevB.82.060518, PhysRevB.84.132503}. In other words, $T^2$ behavior appears in the disordered limit. 
However, the residual resistivity of approximately $50~\mathrm{\mu\Omega cm}$ in the SrFe$_2$(As$_{0.7}$P$_{0.3}$)$_2$ is considerably smaller than that in other iron-based superconductors where $T^2$ behavior has been observed, for example in Ni-doped and Co-doped compounds~\cite{PhysRevB.82.060518,PhysRevB.79.100506}.
This indicates that the exponent $n\simeq 1.6$ is induced by the intrinsic nature of the SC gap structure and not by impurity scattering.
In other P-substituted systems, the observed exponents are very close to unity  ($n<1.2$) and strongly suggest the presence of line nodes.
An exponent $n\simeq1.6$, which is somewhat larger than the exponent values in other systems, indicates that there is another channel of low-energy quasiparticle excitations in addition to the SC gap with line nodes. 

\begin{figure}[tb]
\begin{center}
\includegraphics[width=8.7cm]{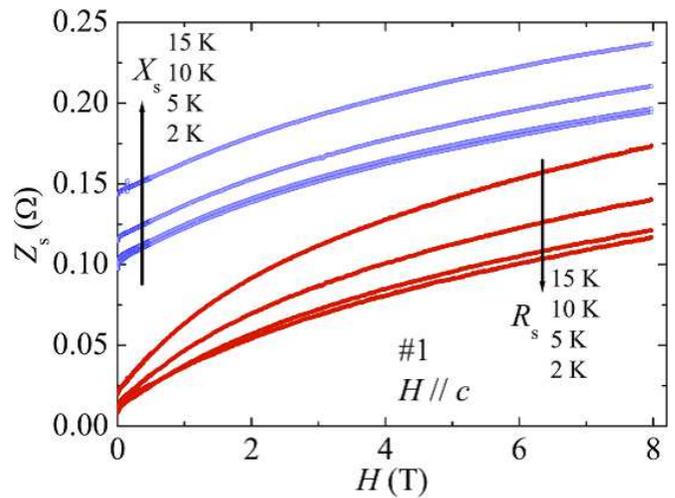}
\caption{(color online) The magnetic field dependence of the surface resistance, $R_s$, and the surface reactance $Xs$ at 44~$\mathrm{GHz}$ under zero-field cooled conditions. }
\label{ZsunderH}
\end{center}
\vspace{-6mm}
\end{figure}

%%%%%%%%%%%%%%%%%%%%%%%%%%%%%
%%%%%%%'±'±'©'玥ê%%%%%%%%%%%%%%%
%%%%%%%%%%%%%%%%%%%%%%%%%%%%%
%\section{Flux flow resistivity}
%%%%%%%%%%%CCmodel'̏Љî%%%%%%%%%%%%%%%%%%%%%
To obtain further insights into the gap structure, we measured  $Z_s$ under finite magnetic fields and extracted the flux flow resistivity, $\rho_f$.
Figure~\ref{ZsunderH} shows the magnetic field dependence of $Z_s$ at different temperatures.
We performed experiments under both field cooled and zero-field cooled conditions and confirmed that there were no significant effects of vortex pinning on $Z_s$ in a wide $H$ range, except for very low fields below 0.5~$\mathrm{T}$ where a small degree of hysteresis appeared. 
In other words, the vortices are distributed homogeneously in the samples.
We used the Coffey-Clem model to analyze $Z_s$~\cite{PhysRevLett.67.386}. In this model, energy dissipation is generated by the motion of a single vortex driven by an electric current, $\bm{J}$, which is described as
\begin{equation}
\eta\dot{\bm{u}}+\kappa_p \bm{u}=\phi_0 \bm{J}\times n,
\end{equation}
where $\bm{u}$ is the displacement of the vortex from its equilibrium pinning site, $\eta$ is the viscous drag coefficient,
and $\kappa_p$ is the restoring force constant.
Surface impedance in the mixed state ($H\gg H_{c1}$) can be expressed as
\begin{equation}
Z_s=-i\mu\omega\sqrt{\frac{\lambda_L^2+(i/2)\tilde{\delta^2_{f}}(1-i\omega_p/\omega)^{-1}}{1+is}},
\end{equation}
where $\tilde{\delta_{f}}=\sqrt{2\rho_f/\mu\omega}=\sqrt{2H \phi_0/\omega\eta}$ is the flux flow skin depth and $s=2\lambda_L^2/\delta_{nf}^2$ denotes the correction term of the normal fluid, which can be neglected in the low-temperature limit by setting $s=0$. 
From this equation, we calculated the pinning frequency, $\omega_p$, and $\rho_f$.
%%%%%%%%%%%%%%%%%%%%%%%%%%%%%%%%%%%%%%%%%

%%%%%%%%%%%%%%%%%%%%%%%%%%%%%%%
%%%%%%%%%%FIgure
%%%%%%%%%%%%%%%%%%%%%%%%%%%%%%%
%%%%%%%%%%%%%%%%%%%%%%%%%%%%%%%

\begin{figure}[tb]
\begin{center}
\includegraphics[width=8.7cm]{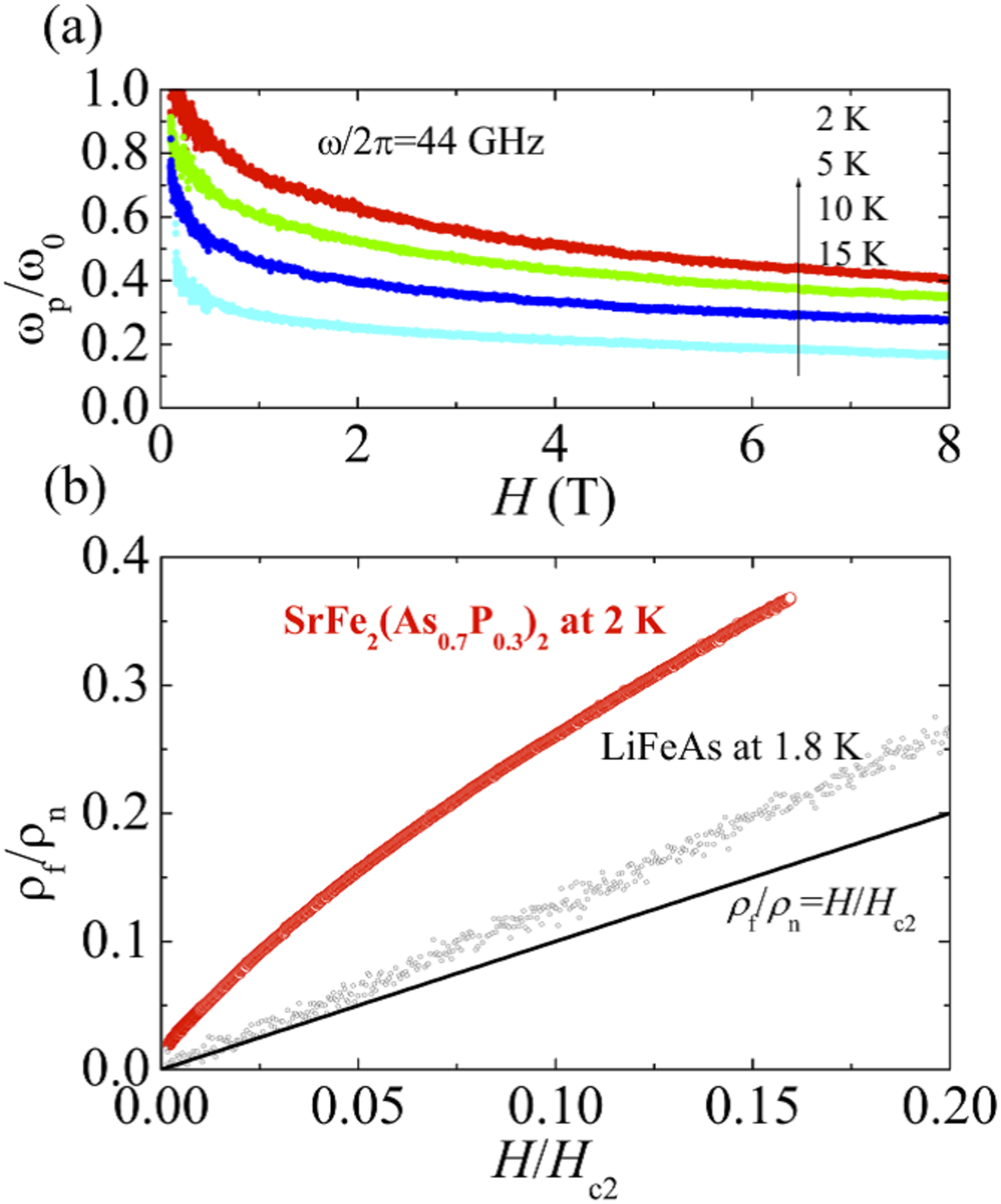}
\caption{(color online) (a)The pinning frequency of SrFe$_2$(As$_{0.7}$P$_{0.3}$)$_2$. (b) The magnetic field dependences of the flux flow resistivities of SrFe$_2$(As$_{0.7}$P$_{0.3}$)$_2$ and LiFeAs~\cite{arXiv:1110.6575}. The solid line indicates the BS relation, $\rho_f/\rho_n=H/H_{c2}$. }
\label{rhof}
\end{center}
\vspace{-6mm}
\end{figure}

Figure~\ref{rhof} (a) illustrates the $\omega_p$ of SrFe$_2$(As$_{0.7}$P$_{0.3}$)$_2$. $\omega_p$ increases at low temperatures and at low fields, and it exceeds $30\ \mathrm{GHz}$ at $2\ \mathrm{K}$ and $1\ \mathrm{T}$, which is considerably larger than the values of $6\ \mathrm{GHz}$ for LaFeAsO$_{0.9}$F$_{0.1}$ \cite{PhysRevB.78.012507} and $3\ \mathrm{GHz}$ for LiFeAs~\cite{arXiv:1110.6575}.
We can also estimate the vortex viscous drag coefficient $\eta$.
This parameter is related to the interval of discrete core bound states, $\omega_c\sim\Delta^2 /E_F$ via $\eta=\pi\hbar n \omega_c\tau$, where $\tau$ is the quasiprticle relaxation time in the vortex core.
$\omega_c\tau$ was observed to be on the order of $10^{-1}$, which indicates that the vortex core is in the moderately clean regime~\cite{PhysRevB.63.184517}.

Figure~\ref{rhof} (b) shows the magnetic field dependence of $\rho_f$. 
It is clear that $\rho_f$ exhibits a steeper increase at lower fields ($\rho_f/\rho_n > 2.5H/H_{c2}$ at $0.1H_{c2}$, where $\rho_n$ is the resistivity in the normal state and $H_{c2}$ is the upper critical field) than it does in the Bardeen and Stephen's (BS) model~\cite{PhysRev.140.A1197}, $\rho_f/\rho_n=H/H_{c2}$, and in the nodeless superconductor LiFeAs ($\rho_f/\rho_n\sim 1.3H/H_{c2}$)~\cite{JPSJ.80.013704,arXiv:1110.6575}.
According to the Kopnin and Volovik (KV) theory~\cite{PhysRevLett.79.1377}, in the moderately clean regime, $\rho_f$ can be expressed as
\begin{equation}
\frac{\rho_f}{\rho_n}=\frac{\omega_c^{\mathrm{max}}}{\Braket{\omega_c (\bm{k})}_{FS}}\frac{H}{H_{c2}},
\end{equation}
where $\Braket{\cdots}_{FS}$ denotes an average over Fermi surface. For an $s$-wave superconductor, this equation is reduced to the BS relation. However, when the SC gap has nodes,  $\omega_c^{\mathrm{max}}/\Braket{\omega_c (\bm{k})}_{FS}$ becomes larger than unity because $\omega_c$ depends on $\bm{k}$, which results in the enhancement of $\rho_f$. 
For example,  $\rho_f/\rho_n\sim 2H/H_{c2}$ is observed in single-gap superconductors with line nodes~\cite{PhysRevB.66.014527}.
Note that it is important to correctly evaluate $H_{c2}(0\ \mathrm{K})$; otherwise, ambiguity arises in the determination of the slope of $\rho_f (H)$.
The $H_{c2}$ of SrFe$_2$(As$_{0.7}$P$_{0.3}$)$_2$ has been measured up to $H_{c2}(8 \ \mathrm{K})=45\ \mathrm{T}$~\cite{Kida}, which results in $H_{c2}(0\ \mathrm{K})=50\ \mathrm{T}$ with the aid of Ginzburg-Landau theory.
Because the $H_{c2} (T)$ of iron-based superconductors often exhibits a strong $T$-dependence even at low temperatures, $H_{c2}(0\ \mathrm{K})$ may be greater than $50~\mathrm{T}$. In that case, the slope of $\rho_f/\rho_n$ will be larger than that in our plot.
Therefore, our plot provides the lower limit of the possible slope.
Even if $H_{c2}$ is greaater than $50~\mathrm{T}$, we can safely state that $\rho_f/\rho_n > 2.5H/H_{c2}$ at $0.1H_{c2}$.

%%%%%%%%%%%%%%%%%%%%%%%%%%%%%%%
%%%%%%%%%%FIgure
%%%%%%%%%%%%%%%%%%%%%%%%%%%%%%%
%%%%%%%%%%%%%%%%%%%%%%%%%%%%%%%

We would like to note two unusual features of $\rho_f$, which cannot be simply taken as the results of the presence of the line nodes confirmed by $\lambda_L (T)$. One is that the enhancement is considerably larger than that in the single-gap superconductor with line nodes. 
In the simplified phenomenological model of multi-gapped superconductors, quasiparticles excited at different Fermi surfaces contribute to the flux flow resistivity, similar to a parallel circuit, $\rho_f^{-1}=\rho_{f,1}^{-1}+\rho_{f,2}^{-1}+\cdots$~\cite{JPSJ.74.1394}. 
Therefore, even if a nodal gap exists, the $\rho_f$ associated with nodeless gaps should suppress the enhancement of the total $\rho_f$.
The other unusual feature is the nonlinear $H$ dependence.
In both KV theory and the theory in Ref.~33, $\rho_f$ is linear in $H$ at low fields.
In addition, we observed that $\rho_f \propto H$ in LiFeAs as shown in Fig.~\ref{rhof}~(b)~\cite{arXiv:1110.6575}.
Nonlinear $H$ dependence and a steep initial gradient are also found in $\rho_f$ for MgB$_2$.
MgB$_2$ has two distinct gaps where the gap amplitude ratio of the larger gap to the smaller gap is $\Delta_{L}/\Delta_{S}\sim 3$.
Because the smaller gap is more sensitive to $H$, quasiparticles excited in the smaller gap rapidly increase with increasing $H$ and enhance the flux flow resistivity at low fields~\cite{PhysRevB.68.060501}. 
Therefore, the contribution from the smaller gap saturates even far below $H_{c2}$. 
In such a situation, $\rho_f$ may not be linear in $H$. 
Because SrFe$_2$(As$_{0.7}$P$_{0.3}$)$_2$ is also a multigap superconductor, it is natural to conclude that the contribution from the very small gaps mentioned above produces unusual features.
However, note that we cannot distinguish between the contribution from small gaps opening over an entire Fermi sheet and that from a deep gap minimum on a limited part of a Fermi sheet.

We now discuss the difference between SrFe$_2$(As$_{0.7}$P$_{0.3}$)$_2$ and BaFe$_2$(As$_{0.67}$P$_{0.33}$)$_2$ in terms of the low-energy quasiparticle excitation.
The low-temperature $\lambda_L(T)$ exhibits different behaviors between these compounds: while BaFe$_2$(As$_{0.67}$P$_{0.33}$)$_2$ exhibits nearly $T$-linear $\lambda_L(T)$~\cite{PhysRevB.81.220501}, we observed that $\lambda_L (T)\propto T^{1.6}$ in SrFe$_2$(As$_{0.7}$P$_{0.3}$)$_2$.
As we already discussed, both behaviors are consistent with the presence of a SC gap with line nodes.
Therefore, we consider that the origin of the difference between these two compounds is related to the differences in the structure of the nodeless gaps.
The significant differences in the low-energy quasiparticle excitation between these two compounds are also observed in the spin-relaxation rate and the heat capacity~\cite{PhysRevB.85.144515}. 
Both quantities exhibit a considerably larger residual density of states under magnetic fields in SrFe$_2$(As$_{1-x}$P$_{x}$)$_2$, which is consistent with the presence of the very small gaps that we observed in $\rho_f$.
In the Sr122 system, additional quasiparticle excitations appear by replacing Ba$^{2+}$ with Sr$^{2+}$.
Because the substitution of Ba$^{2+}$ by Sr$^{2+}$ introduces enhanced three-dimensionality, we can conclude that additional quasiparticle excitations observed in the above experiments are related to the strong modulation of the SC gap structure along the $k_z$ direction rather than small gaps opening over a whole Fermi sheet.
We also believe that the fractional power of $\lambda_L(T)$ is due to the combination of quasiparticle excitations in the nodal gap and the deep minimum in the nodeless gaps.
In fact, Mishra \textit{et al.}~\cite{PhysRevB.84.014524} calculated the magnetic penetration depth assuming the presence of three-dimensional nodes or horizontal nodes and found that $n$ can be 1.5-2 when there is a nodeless gap with deep minimum on other Fermi surfaces. 

In conclusion, we measured the microwave surface impedances of SrFe$_2$(As$_{0.7}$P$_{0.3}$)$_2$ single crystals. The superfluid density at low temperatures obeys a power law, $n_s (T)/n_s (0)=1-C(T/T_c)^n$, with a fractional exponent of $n=1.5$-1.6. The flux flow resistivity exhibits a very sharp increase with increasing $H$ and a nonlinear field dependence. These results indicate the presences of both a gap with line nodes and nodeless gaps with a deep minimum.
We argued that the details of the SC gap structure depend on the size of the alkaline earth ion between the Fe-As layers in the 122 system. 
However, despite changing the details of the SC gap structure, the Sr122 system maintains a $T_c$ comparable to that in the Ba122 system.
This result indicates that the change caused by replacing Ba$^{2+}$ with Sr$^{2+}$ occurs in the region of the Fermi surface that does not significantly affect the pairing interaction in these compounds. 
Therefore, by comparing these two compounds in more detail, we will be able to clarify the mechanism of iron-based superconductivity.

\providecommand{\noopsort}[1]{}\providecommand{\singleletter}[1]{#1}

\end{document}